# Learning Hybrid Algorithms for Vehicle Routing Problems

Yves Caseau[1], Glenn Silverstein[2] , François Laburthe[1]

[1] BOUYGUES e-Lab., 1 av. E. Freyssinet, 78061 St. Quentin en Yvelines cedex, FRANCE
ycs;flaburth@bouygues.com
[2] Telcordia Technologies, 445 South Street, Morristown, NJ, 07960, USA
silverst@research.telcordia.com

**Abstract**

This paper presents a generic technique for improving hybrid algorithms through the discovery of and tuning of meta-heuristics. The idea is to represent a family of "push/pull" heuristics that are based upon inserting and removing tasks in a current solution, with an algebra. We then let a learning algorithm search for the best possible algebraic term, which represents a hybrid algorithm for a given set of problems and an optimization criterion. In a previous paper, we described this algebra in detail and provided a set of preliminary results demonstrating the utility of this approach, using vehicle routing with time windows (VRPTW) as a domain example. In this paper we expand upon our results providing a more robust experimental framework and learning algorithms, and report on some new results using the standard Solomon benchmarks. In particular, we show that our learning algorithm is able to achieve results similar to the best-published algorithms using only a fraction of the CPU time. We also show that the automatic tuning of the best hybrid combination of such techniques yields a better solution than hand tuning, with considerably less effort.

# 1. Introduction

Recent years have seen a rise in the use of hybrid algorithms in many fields such as scheduling and routing, as well as generic techniques that seem to prove useful in many different domains (e.g., Limited Discrepancy Search (LDS) [HG95] and Large Neighborhood Search (LNS) [Sha98]). Hybrid Algorithms are combinatorial optimization algorithms that incorporate different types of techniques to produce higher quality solutions. Although hybrid algorithms and approaches have achieved many interesting results, they are not a panacea yet as they generally require a large amount of tuning and are often not robust enough: i.e., a combination that works well for a given data set does poorly on another one. In addition, the application to the "real world" of an algorithm that works well on academic benchmarks is often a challenging task.

The field of Vehicle Routing is an interesting example. Many real-world applications rely primarily on insertion algorithms, which are known to be poor heuristics, but have two major advantages: they are incremental by nature and they can easily support the addition of domain-dependent side constraints, which can be utilized by a constraint solver to produce a high-quality insertion [CL99]. We can abstract the routing aspect and suppose that we are solving a multi-resource scheduling problem, and that we know how to incrementally insert a new task into a current solution/schedule, or how to remove one of the current tasks. A simple approach is to derive a greedy heuristic (insert all the tasks, using a relevant order); a simple optimization loop is then to try 2-opt moves where pairs of tasks (one in and one out) are swapped. The goal of our work is to propose a method to (1) build more sophisticated hybrid strategies that use the same two push and pull operations; (2) automatically tune the hybrid combination of meta-heuristics. In [CSL99] we described such a framework for discovering and tuning hybrid algorithms for optimization problems such as vehicle routing based on a library of problem independent meta-methods. In this paper, we expand upon the results in [CLS99] providing a more robust experimentation framework and experimentation methodology with more emphasis on the automated learning and tuning of new terms.

This paper is organized as follows. Section 2 presents a generic framework that we propose for some optimization problems, based on the separation between domain-specific low-level methods, for which constraint solving is ideally suited, and generic meta-heuristics. We then recall the definition of a Vehicle Routing Problem and show what type of generic meta-heuristics may be applied. Section 3 provides an overview of the algebra of terms representing the hybrid methods. Section 4 describes a learning algorithm for inventing and tuning terms representing problem solutions. Section 5 describes an experimentation framework, which is aimed at discovering the relative importance of various learning techniques (mutation, crossover, and invention) and how they, along with experiment parameters such as the pool size and number of iterations, affect the convergence rate and stability. Sections 6 and 7 provide the results of a various experiments along with the conclusions.

## 2. Meta-Heuristics and Vehicle Routing

### 2.1 A Framework for Problem Solving

The principle of this work is to build a framework that produces efficient problem solving algorithms for a large class of problems at reduced development cost, while using state-of-the-art meta-heuristics. The main idea is the separation between two parts, as explained in Figure 1, that respectively contain a domain-dependent implementation of two simple operations (push and pull) and contain a set of meta-heuristics and a combination engine, that are far more sophisticated but totally generic. Obviously, the first postulate is that many problems may actually fit this scheme. This postulate is based on our experience with a large number of industrial problems, where we have found ourselves to re-using "the same set of tricks" with surprisingly few differences. Here is a list of such problems:

(1) *Vehicle Routing Problems*. As we shall later see, such problems come in all kinds of flavor, depending on the objective function (what to minimize) and the side-constraints (on the trucks, on the clients, etc.). The key methods are the insertion of a new task into a route, which is precisely the resolution of a small TSP with side constraints.

(2) *Assignment Problems*. We have worked on different types of assignment problems, such as broadcast airspace optimization or workflow resource optimization that may be seen as assigning tasks to constrained resources. The unit operations are the insertion and the removal of one task into one resource.

(3) *Workforce Scheduling*. Here the goal is to construct a set of schedules, one for each operator, so that a given workload, such as a call center distribution, is handled optimally. The algorithm that has produced the best overall results and that is being used today in our software product is based on decomposing the workload into small units that are assigned to workers. The unit "push" operation is the resolution of a small "one-machine" scheduling problem, which can be very tricky because of labor regulations. Constraint solving is a technique of choice for solving such problems.

(4) *Frequency Allocation*. We participated to the 2001 ROADEF challenge [Roa01] and applied this decomposition to a frequency allocation problem. Here the unit operation is the insertion of a new path into the frequency plan. This problem is solved using a constraint-based approach similar to jobshop scheduling algorithms [CL95].

For all these problems, we have followed a similar route: first, build a greedy approach that is quickly extended into a branch-and-bound scheme, which must be limited because of the size of the problem. Then local-optimization strategies are added, using a swap (push/pull) approach, which are extended towards large neighborhood search methods.

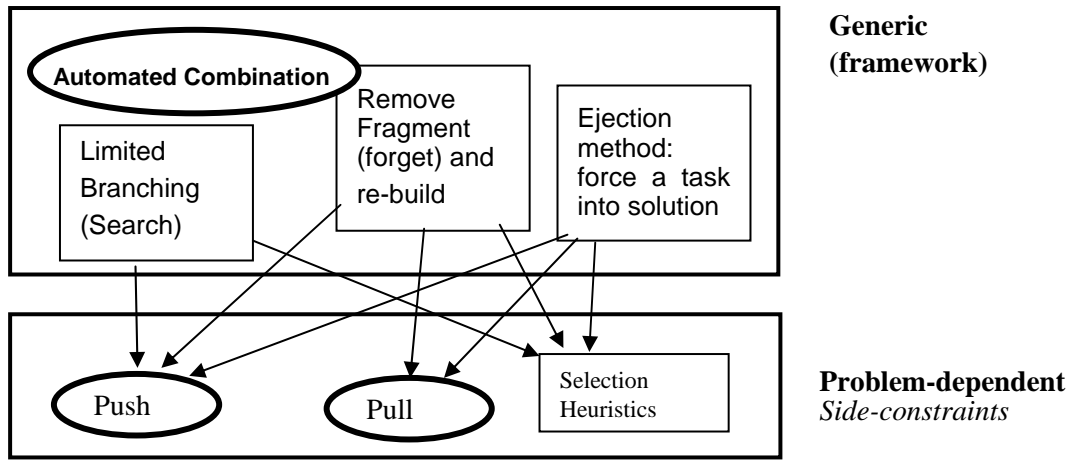

*Figure 1:A Framework for Meta-Heuristics*

The problem that we want to address with this paper is twofold:
1. How can we build a library of meta-heuristics that is totally problem-independent, so that any insertion algorithm based on a constraint solver can be plugged?
2. How can we achieve the necessary tuning to produce at a low cost a robust solution for each different configuration?

The first question is motivated by the fact that the meta-heuristic aspect (especially with local optimization) is the part of the software that is most difficult to maintain when new constraints are added. There is a tremendous value in confining the domain-dependent part to where constraint-programming techniques can be used. The second question is drawn from our practical experience: the speed of the hardware, the runtime constraints, the objective functions (total travel, number of routes, priority respect,...) all have a strong influence when designing a hybrid algorithm. For instance, it is actually a difficult problem to re-tune a hybrid algorithm when faster hardware is installed.

## 2.2 Application to the Vehicle Routing Problem

### 2.2.1 Vehicle Routing Problems

A vehicle routing problem is defined by a set of tasks (or nodes) *i* and a distance matrix (d[*i,j*]). Additionally, each task may be given a duration (in which case the matrix d denotes travel times) and a load, which represents some weight to be picked in a capacitated VRP. The goal is to find a set of routes that start from a given node (called the depot) and return to it, so that each task is visited only once and so that each route obeys some constraints about its maximal length or the maximum load (the sum of the weights of the visited tasks). A VRP is an optimization problem, where the objective is to minimize the sum of the lengths of the routes and/or the number of routes [Lap92].

We report experiments with the Solomon benchmarks [Sol87], which are both relative small (100 customers) and simple (truck capacity and time-windows). Real-world routing problems include a lot of side constraints (not every truck can go everywhere at anytime, drivers have breaks and meals, some tasks have higher priorities, etc.). Because of this additional complexity, the most commonly used algorithm is an insertion algorithm, one of the simplest algorithms for solving a VRP.

Let us briefly describe the VRPTW problem that is proposed in the Solomon benchmarks more formally.

- We have N = 100 customers, defined by a load $c_i$, a duration $d_i$ and a time window $[a_i, b_i]$
- $d(i,j)$ is the distance (in time) to go from the location of customer $i$ to the location of customer $j$ (all trucks travel at the same speed). For convenience, 0 represents the initial location of the depot where the trucks start their routes.
- A solution is given by a route/time assignment: $r_i$ is the truck that will service the customer $i$ and $t_i$ is the time at which this customer will be serviced. We define $prev(i) = j$ if $j$ is the customer that precedes $i$ in the route $r_i$ and $prev(i) = 0$ is $i$ is the first customer in $r_i$. We also define $last(r)$ as the last customer for route $r$.
- The set of constraints that must be satisfied is the following:
  - $\forall i, \ t_i \in [a_i, b_i] \wedge t_i \geq d(0,i)$
  - $\forall i,j \ \ r_i = r_j \Rightarrow t_j - t_i \geq d_i + d(i,j) \vee t_i - t_j \geq d_j + d(j,i)$
  - $\forall r, \ \sum_{i \in \{i | r_i = n\}} c_i \leq C$
- The goal is to minimize E = $\max_{i \leq N}(r_i)$ (the total number of routes) and
  $$D = \sum_{r \leq K} length(r) \ \text{ with } length(r) = \sum_{i | r_i = r} d(prev(i),i) + d(last(r), 0)$$

There are two reasons for using the Solomon benchmarks instead of other larger problems. First, they are the only problems for which there are many published results. Larger problems, including our own combined benchmarks (with up to 1000 nodes [CL99]), did not receive as much attention yet and we will see that it is important to have a good understanding of competitive methods (i.e., quality vs. run-time trade-off) to evaluate how well our learning approach is doing. Second, we have found using large routing problems from the telecommunication maintenance industry that these benchmarks are fairly representative: techniques that produced improvements on the Solomon benchmarks actually showed similar improvement on larger problems, provided that their run-time complexity was not prohibitive. Thus, our goal in this paper is to focus on the efficient resolution of Solomon problems, with algorithms that could later scale up to larger problems (which explains our interest for finding good solutions within a short span of time, from a few seconds to a minute).

### 2.2.2 Insertion and Incremental Local Optimization

Let us first describe an insertion-based greedy algorithm. The tasks (i.e., customers) to be visited are placed in a stack, that may be sorted statically (once) or dynamically, and a set of empty routes is created. For each task, a set of candidate routes is selected and the feasibility of the insertion is evaluated. The task is inserted into the best route found during this evaluation. This loop is run until all tasks have been inserted. Notice that an important parameter is the valuation of the feasible route. A common and effective strategy is to pick the route for which the increase in length due to the insertion is minimal.

The key component is the node insertion procedure (i.e., the push operation of Figure 1), since it must check all the side constraints. CP techniques can be used either through the full resolution of the one-vehicle problem, which is the resolution of a small with side-constraints [CL97][RGP99], or it can be used to supplement a simple insertion heuristic by doing all the side-constraint checking. We have shown in [CL99] that using a CP solver for the node insertion increases the quality of the global algorithm, whether this global algorithm is a simple greedy insertion algorithm or a more complex tree search algorithm.

The first hybridization that we had proposed in [CL99] is the introduction of incremental local optimization. This is a very powerful technique, since we have shown that it is much more efficient than applying local optimization as a post-treatment, and that it scales very well to large problems (many thousands of nodes). The interest of ILO is that it is defined with primitive operations for which the constraint propagation can be easily implemented. Thus, it does not violate the principle of separating the domain-dependent part of the problem from the optimization heuristics.

Instead of applying the local moves once the first solution is built, the principle of incremental local optimization is to apply them after each insertion and only for those moves that involve the new node that got inserted. The idea of incremental local optimization within an insertion algorithm had already brought good results in [GHL94], [Rus95] and [KB95]. We have applied ILO to large routing problems (up to 5000 nodes) as well as call center scheduling problems.

Our ILO algorithm uses three moves, which are all 2- or 3- edge exchanges. The first three are used once the insertion is performed. These moves are performed in the neighborhood of the inserted node, to see if some chains from another route would be better if moved into the same route. They include a 2-edge *exchange* for crossing routes (see Figure 2), a 3-edge exchange for *transferring* a chain from one route to another and a simpler *node transfer* move (a limited version of the chain transfer).

The 2-edge move (i.e., exchange between $(x,y)$ and $(i,i')$) is defined as follows. To perform the exchange, we start a branch where we perform the edge substitution by linking $i$ to $y$ and $i'$ to $x$. We then compute the new length of the route $r$ and check side constraints if any apply. If the move is illegal we backtrack to the previous state. Otherwise, we perform a route optimization on the two modified routes (we apply 3-opt moves within the route $r$). We also recursively continue looking for 2-opt moves and we apply the greedy 3-opt optimization to $r'$ that will be defined later.

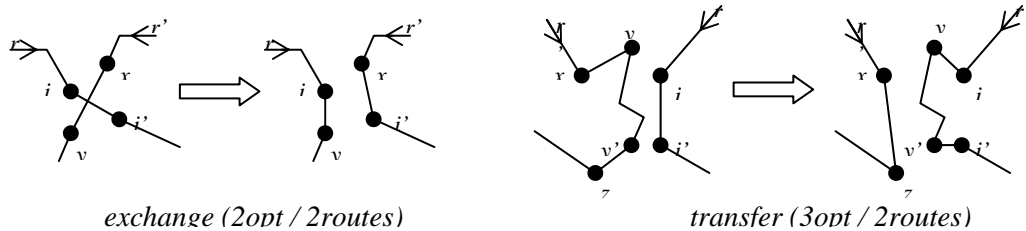

*exchange (2opt / 2routes)*    *transfer (3opt / 2routes)*

*Figure 2: edge exchange moves used for ILO*

The second move transfers a chain ($y \to y'$) from a route $r'$ to a route $r$ right after the node $i$. We use the same implementation technique and create a branch where we perform the 3-edge exchange by linking $i$ to $y$, $y'$ to $i'$ and the predecessor $x$ of $y$ to the successor $z$ of $y'$. We then optimize the augmented route r (assuming the new route does not violate any constraints) and check that the sum of the lengths of the two resulting routes has decreased. If this is not the case, we backtrack; otherwise, we apply the same optimization procedures to r' as in the precedent routine.

We also use a more limited version of the transfer routine that we can apply to a whole route (as opposed to the neighborhood of a new node i). A « greedy optimization » looks for nodes outside the route r that are close to one node of the route and that could be inserted with a gain in the total. The algorithm is similar to the previous one, except that we do not look for a chain to transfer, but simply a node.

In the rest of the paper, we will call INSERT(*i*) the insertion heuristic obtained by applying greedily the node insertion procedure and a given level of ILO depending on the value i:

$i = 0 \Leftrightarrow$ no ILO

$i = 1 \Leftrightarrow$ perform only 2-opt moves (*exchange*)

*i = 2* ⇔ performs 2 and 3-opt moves (*exchange* and *transfer*)

*i = 3* ⇔ perform 2 and 3-opt moves, plus greedy optimization

*i = 4* ⇔ similar, but in addition, when the insertion fail we try to reconstruct the route by inserting the new node first.

## 2.3 Meta-Heuristics for Insertion Algorithms

In the rest of this section we present a set of well-known meta-heuristics that have in common the property that they only rely on inserting and removing nodes from routes. Thus, if we can use a node insertion procedure that checks all domain-dependent side-constraints, we can apply these meta-heuristics freely. It is important to note that not all meta-heuristics have this property (e.g., splitting and recombining routes does not), but the ones that do make a large subset and we will show that these heuristics, together with ILO, yield powerful hybrid combinations.

### 2.3.1 Limited Discrepancy Search

Limited Discrepancy Search is an efficient technique that has been used for many different problems. In this paper, we use the term LDS loosely to describe the following idea: transform a greedy heuristic into a search algorithm by branching only in a few (i.e., limited number) cases when the heuristic is not "sure" about the best insertion. A classical complete search (i.e., trying recursively all insertions for all nodes) is impossible because of the size of the problem and a truncated search (i.e., limited number of backtracks) yields poor improvements. The beauty of LDS is to focus the "power of branching" to those nodes for which the heuristic decision is the least compelling. Here the choice heuristic is to pick the feasible route for which the increase in travel is minimal. Applying the idea of LDS, we branch when two routes have very similar "insertion costs" and pick the obvious choice when one route clearly dominates the others. There are two parameters in our LDS scheme: the maximum number of branching points along a path in the search tree and the threshold for branching. A low threshold will provoke a lot of branching in the earlier part of the search process, whereas a high threshold will move the branching points further down. These two parameters control the shape of the search tree and have a definite impact on the quality of the solutions.

### 2.3.2 Ejection Chains and Trees

The search for **ejection chains** is a technique that was proposed a few years ago for Tabu search approaches [RR96]. An ejection link is an edge between *a* and *b* that represents the fact that *a* can be inserted in the route that contains *b* if *b* is removed. An ejection chain is a chain of ejection edges where the last node is free, which means that it can be inserted freely in a route that does not intersect the ejection chain, without removing any other node. Each time an ejection chain is found, we can compute its cost, which is the difference in total length once all the substitutions have been performed (which also implies the insertion of the root node).

The implementation is based on a breadth-first search algorithm that explores the set of chains starting from a root *x*. We use a marker for each node *n* to recall the cost of the cheapest ejection chain that was found from *x* to *n*, and a reverse pointer to the parent in the chain. The search of ejection chains was found to be an efficient technique in [CL98] to minimize the number of routes by calling it each time no feasible insertion was found during the greedy insertion. However, it is problem-dependent since it only works well when nodes are of similar importance (as in the Solomon benchmarks). When nodes have different processing times and characteristics, one must move to ejection trees.

An **ejection tree** is similar to an ejection chain but we allow multiple edges from one node *a* to $b_1, .. , b_n$ to represent the fact that the "forced insertion" of *a* into a route *r* causes the ejection of $b_1,..,b_n$. For one node *a* and a route *r*, there are usually multiple subsets of such $\{b_1, .., b_n\}$ so we use a heuristic to find a set as small as possible. An ejection tree is then a tree of root *a* such that all leaves are free nodes that can be inserted into different routes that all have an empty intersection with the tree. There are many more ejection trees than there are chains, and the search for ejection trees with lowest possible cost must be controlled with topological parameters (maximum depth, maximum width, etc.). The use of ejection trees is very similar to the use of ejection chains, i.e. we can use it to insert a node that has no feasible route, or as a meta-heuristic by removing and re-inserting nodes.

The implementation is more complex because we cannot recursively enumerate all trees using a marking algorithm. Thus we build a search tree using a stack of ejected nodes. When we start, the stack contains the root node; then each step can be described as follows:

- Pick a node *n* from the stack.
- For all routes *r* into which a forced insertion is possible, create a branch of the search tree. In this branch, perform the insertion (*a* into *r*) and produce a subset of nodes that got "ejected". To keep that set as small as possible, every free node is immediately re-inserted. The set of ejected nodes is then placed into the stack.
- If the stack is empty we register the value of the current tree (each node in the search tree corresponds to an ejection tree)

To make this algorithm work, it is necessary to put an upper bound on the depth of the tree and on the branching factor. We only select no more than *k* routes for the forced insertion, by filtering the *k* best routes once all possible routes have been tried. Furthermore, we use a LDS scheme: each time we use a route that was not the best route found (the valuation is simply the weight of the ejected set), we count one discrepancy. The LDS approach simply means to cut all trees that would require more than *D* (a fixed parameter) discrepancies.

### 2.3.3 Large Neighborhood Search

Large Neighborhood Search (LNS) is the name given by Shaw [Sha98] to the application of shuffling [CL95] to routing. The principle is to forget (remove) a fragment of the current solution and to rebuild it using a limited search algorithm. For jobshop scheduling, we have developed a large variety of heuristics to determine the fragment that is forgotten and we use them in rotation until a fix-point is reached. We have used a truncated search to re-build the solution, using the strength of branch-and-bound algorithms developed for jobshop scheduling. In his paper, Shaw introduced a heuristic randomized criterion for computing the "forgotten" set and proposed to use LDS to re-build the solution. Since he obtained excellent results with this approach, we have implemented the same heuristic to select the set of *n* (an integer parameter) nodes that are removed from the current solution. His procedure is based on a relatedness criteria and a pseudo-random selection of successive "neighbors". A parameter is used to vary the heuristic from deterministic to totally random. We have extended this heuristic so that nodes that are already without a route are picked first (when they exist).

The implementation of LNS is then straightforward: select a set of *k* nodes using Shaw's procedure and then remove them from the current solution. These nodes are then re-inserted using a LDS insertion algorithm. There are, therefore, four parameters needed to describe this algorithm: two for LDS (number of discrepancies and threshold), the randomness parameter and the number of nodes to be reinserted. As we shall later see, the procedure for re-constructing the solution could be anything, which opens many possible combinations. Notice that the heuristic for selecting the fragment to remove is clearly problem-dependent. For VRP, we use Shaw's technique, for frequency allocation, we had to come up with a fairly complex new method that computes the number of constraint violations for all tasks that could not be inserted.

## 3. An Algebra of Hybrid Algorithms

### 3.1 Representing the Combination of Meta-Heuristics

We represent hybrid algorithms obtained through the composition of meta-heuristics with algebraic formulas (terms). As for any term algebra, the grammar of all possible terms is derived from a fixed set of operators, each of them representing one of the resolution/optimization techniques that we presented in the previous section. There are two kinds of terms in the grammar: <Build> terms represent algorithms that create a solution (starting from an empty set of routes) and <Optimize> terms for algorithms that improve a solution (we replace a set of routes with another one). A typical algorithm is, therefore, the composition of one <Build> term and many <Optimize> terms.

More precisely, we may say that a *build* algorithm has no parameter and returns a solution object that represents a set of routes. Each route is defined as a linked list of nodes. An *optimize* algorithm has one input parameter, which is a current solution and returns another set of routes. In addition, a global object is used to represent the optimization context, which tells which valuation function should be used depending on the optimization criterion.

```
<Build>::    INSERT(i)   |
             <LDS> |
             DO( <Build>,<Optimize>)   |
             FORALL(<LDS>, <Optimize>)

<Optimize> ::   CHAIN(n,m)   |
                TREE(n,m,k)   |
                LNS(n,h,<Build>)   |
                LOOP(n,<Optimize>)   |
                THEN(<Optimize>, …,

<LDS> ::  LDS(i,n,l)
```

*Figure 3: A grammar for hybrid algorithms*

The definition of the elementary operators is straightforward:
- INSERT($i$) builds a solution by applying a greedy insertion approach and a varying level of ILO according to the parameter $i$ (cf. Section 2.2, 0 means no ILO and 4 means full ILO)
- LDS($i,n,l$) builds a solution by applying a limited discrepancy search on top of the INSERT($i$) greedy heuristic. The parameter $n$ represents the maximum number of discrepancies (number of branching points for one solution) and $l$ represents the threshold. A LDS term can also be used as a generator of different solutions when it is used in a FORALL.
- FORALL($t1, t2$) produces all the solutions that can be built with $t1$, which is necessarily a LDS (as opposed to only the best) and applies the post-optimization step $t2$ to each of them. The result is the best solution that was found.
- CHAIN($n,m$) is a post-optimization step that select n nodes using the heuristic represented by m and successively removes them (one at a time) and tries to re-insert them using an ejection chain. We did not come up with any significant selection heuristic so we mostly use the one presented in Section 2.3.2.
- TREE($n,m,k$) is similar but uses an ejection tree strategy for the post-optimization. The extra-parameter represents the number of discrepancies for the LDS search (of the ejection tree).
- LNS($n,h,t$) applies Large Neighborhood Search as a post-optimization step. We select n nodes using Shaws's heuristics with the $h$ randomness parameter and we rebuild the solution using the algorithm represented by $t$, which must be a <Build>. Notice that we do not restrict ourselves to a simple LDS term.
- DO($t1,t2$) simply applies $t1$ to build a solution and $t2$ to post-optimize it
- THEN($t1,t2$) is the composition of two optimization algorithms $t1$ and $t2$

- LOOP(*n,t*) repeats *n* times the optimization algorithm *t*. This is used with an optimization algorithm, which repetition will incrementally improve the value of the current solution.

Here are some examples of algebraic terms.

*LDS(3,3,100)* represents a LDS search using the 3$^{rd}$ level of ILO (every move is tried) but with the regular insertion procedure, trying $2^3$ solutions (3 choice points when the difference between the two best routes is less than 100) and returning the best.

*DO(INSERT(2),CHAIN(80,2))* is an algorithm obtained by combining a regular greedy heuristic with the 2$^{nd}$ level of ILO with a post-optimization phase of 80 removal/re-insertion through an ejection chain.

*FORALL(LDS(0,4,100),LOOP(3,TREE(5,2)))* is an algorithm that performs a LDS search with no ILO and $2^4$ branching points and then applies 3 times an ejection tree post-optimization step for each intermediate solution.

## 3.2 Evaluation

To evaluate the algorithms represented by the terms, we have defined a small interpreter to apply the algorithm represented by the operator (a <Build> or an <Optimize>) to the current problem (and solution for an <Optimize>). The metric for complexity that we use is the number of calls to the insertion procedure. This is a reasonable metric since the CPU time is roughly linear in the number of insertions and has the advantage that it is machine independent and is easier to predict based on the structure of the term. To evaluate the quality of a term, we run it on a set of test files and average the results. The generic objective function is defined as the sum of the total lengths plus a penalty for the excess in the number of routes over a pre-defined objective. In the rest of the paper, we report the number of insertions and the average value of the objective function. When it is relevant (in order to compare with other approaches) we will translate them into CPU (s) and (number of routes, travel).

In [CSL99] we used the algebra to test a number of hand generated terms in the algebra to evaluate the contribution of ILO, and the other four search meta-heuristics comparing algorithms both with and without each of the meta-heuristics. The results are summarized in Table 1, which shows different examples of algebraic terms that represent hybrid combinations of meta-heuristics. For instance, the first two terms (INSERT(3) and INSERT(0)) are a simple comparison of the basic greedy insertion algorithm with and without ILO. In a previous paper [CL99] we had demonstrated that applying 2- and 3-opt moves (with a hill-climbing strategy) as a post-processing step was both much slower and less effective than ILO. Here we tried a different set of post-optimization techniques, using two different objective functions: the number of routes and the total travel time. In this section, the result is the average for the 12 R1* Solomon benchmarks.

We measure the run-time of each algorithm by counting the number of calls to the insertion sub-procedure. The ratio with CPU time is not really constant but this measure is independent of the machine and makes for easier comparisons. For instance, on a PentiumIII-500Mhz, 1000 insertions (first term) translate into 0.08 s of CPU time and 100K (4$^{th}$ term) translates into 8s.

In order to experiment with different optimization goals, we have used the following objective function:

$$f = \frac{1}{|test\_cases|} \times \sum_{t \in test\_cases}[(E \times 1000000 \times \min(1,\max(0,E-E_{opt,t})))+D+\sum_{i \leq N}d_i]$$

Notice that we use $E_{opt,t}$ to represent the optimal number of routes that is known for test case *t*. If we use $E_{opt,t} = 25$, we simply optimize the total travel time for the solution. The constant term (sum of the durations) comes from using this formula to evaluate a partial solution. In the following table, we will use this function to compare different algorithms, but also report the average number of routes when appropriate. Columns 2 and 3 correspond to minimizing the number of trucks, whereas columns 4 and 5 correspond to minimizing the total travel time.

| Term *objective* | Value *# of trucks* | #of insertions *# of trucks* | Value *travel* | # of insertions *travel* |
|---|---|---|---|---|
| INSERT(0) | 14692316 | 1000 | 25293 | 1000 |
| INSERT(3) | 13939595 | 1151 | 22703 | 1167 |
| LDS(3,8,50) | 49789 (12.83) | 162K | 22314 | 182K |
| DO(LDS(4,3,100),CHAIN(80,2)) | 52652 (12.91) | 100K | 22200 | 100K |
| DO(LDS(3,3,100), LOOP(30,LNS(10,4,LDS(4,4,1000)) | 41269 (12.66) | 99K | 22182 | 104K |
| FORALL(LDS(3,2,100),CHAIN(20,2)) | 54466 | 101K | 22190 | 100K |
| DO(INSERT(3),CHAIN(90,2)) | 10855525 | 97k | 22219 | 98K |
| DO(LDS(3,2,100),LOOP(2,TREE(40,2))) | 43564 | 90K | 22393 | 63K |
| DO(LDS(3,2,100),LOOP(6,CHAIN(25,2))) | 56010 | 101K | 22154 | 108K |
| *SuccLNS* = DO(LDS(3,0,100), THEN(LOOP(50,LNS(4,4,LDS(3,3,1000))), LOOP(40,LNS(6,4,LDS(3,3,1000))), LOOP(30,LNS(8,4,LDS(3,3,1000))), LOOP(20,LNS(10,4,LDS(3,3,1000))), LOOP(10,LNS(12,4,LDS(3,3,1000))) )) | 40181 | 26K | 22066 | 50K |

*Table 1. A few hand-generated terms*

In these preliminary experiments, LNS dominated as the technique of choice (coupled with LDS and ejection tree optimization works well on larger problems. Additional experiments in [CSL99] showed that the combination of these techniques with ILO worked better than LNS alone. We can also notice here that the different heuristics are more-or-less suited for different objective functions, which will become more obvious in the next section. We use the name "succLNS" to represent the last (large) term, which is the best from this table and was generated as a (simplified) representation of the strategy proposed in [Sha98], with the additional benefit of ILO.

## 4. A Learning Algorithm for discovering New Terms

### 4.1 Tools for learning

The primary tools for learning are invention of new terms, mutation of existing terms, and the crossing of existing terms with each other. Throughout learning a pool of the best n terms is maintained (the pool size n is a fixed parameter of the experiment) from which new terms are added and created from existing terms. The **invention** of new terms is defined by structural induction from the grammar definition. The choice among the different subclasses (e.g. what to pick when we need an <Optimize>) and the values for their parameters are made using a random number generator and a pre-determined distribution. The result is that we can create terms with an arbitrary complexity (there are no boundaries on the level of recursion, but the invention algorithm terminates with probability 1). One of the key experimental parameters, which are used to guide invention, is a bound on the complexity for the term (i.e., the complexity goal). The complexity of a term can be estimated from its structure and only those terms that satisfy the complexity goal will be allowed to participate in the pool of terms.

Complexity is an estimate of the number of insertions that will be made when running one of the hybrid algorithms. For some terms, this number can be computed exactly, for others it is difficult to evaluate precisely. However, since we use complexity as a guide when inventing a new term, a crude estimate is usually enough. Here is the definition that we used.

- complexity(INSERT($i$)) = 1000.
- complexity(LDS($i,n,l$)) = (if ($i$ = 4) 6000 else 1000) × ($2^n$).
- complexity(FORALL($t1, t2$)) = complexity($t_1$) + complexity($t_2$)
- complexity(CHAIN($n,m$)) = 1500 × $n$).
- complexity(TREE($n,m,k$)) = 600 × $n$ × ($2^k$).
- complexity(LNS($n,h,t$)) = (complexity($t$)$^n$) / 100,.
- complexity(DO($t_1,t_2$)) = complexity($t_1$) + complexity($t_2$)
- complexity(THEN($t_1,t_2$)) = complexity($t_1$) + complexity($t_2$)
- complexity(LOOP($n,t$)) = $n$ ×complexity($t$).

**Mutation** is also defined by structural induction according to two parameters. The first parameter tells if we want a shallow modification, an average modification or a deep modification. In the first case, the structure of the term does not change and the mutation only changes the leaf constants that are involved in the term (integers). Moreover, only small changes for these parameters are supported. In the second case, the type (class) does not change, but large changes in the parameters are allowed and, with a given probability and some sub-terms can be replaced by terms of other classes. In the last case, a complete substitution with a different term is allowed (with a given probability). The second parameter gives the actual performance of the term, as measured in the experiment. The mutation algorithm tries to adjust the term in such a way that the (real) complexity of the new term is as close as possible to the global objective. This compensates the imprecision of the complexity estimate quite effectively. The definition of the mutation operator is a key component of this approach. If mutation is too timid, the algorithm is quickly stuck into local optimums. The improvements shown in this paper compared to our earlier work [CSL99] are largely due to a better tuning of the mutation operator. For instance, although we guide the mutation according to the observed complexity (trying to raise or lower it), we randomly decide (10% of the time) to ignore this guiding indication.

The complete description of the mutation operator is too long to be given in this paper, but the following is the CLAIRE [CL96] method that we use to mutate a AND term. The two parameters are respectively a Boolean that tells if the current term x is larger or smaller than the complexity goal and an integer (i) which is the previously mentioned mutation level. The AND object to which mutation is applied (x = AND(t1,t2)) has two slots, x.optim = t1 and x.post = t2. We can notice that the default strategy is simply to recursively apply mutation to t1 and t2, but we randomly select more aggressive strategies either to simplify the term or to get a more complex one.

```
mutate(x:AND,s:boolean,i:integer) : Term
  -> let y := random(100 / i), y2 := random(100) in
       (if ((s & y2 > 20) | y > 90)        // try to get a more complex term
          (if (y < 10)   THEN(x,invent(Optimizer))
           else if (y < 20)THEN(optim(x),LOOP(randomIn(3,10), post(x)))
           else if (y < 30)THEN(LOOP(randomIn(3,10),optim(x)), post(x))
           else  // recursive application of mutation
                THEN(mutate(optim(x),s,i), mutate(post(x),s,i)))
        else                              // try to get a simpler term
          (if ((i = 3 & y < 50) | (i > 1 & y < 10)) optim(x)
           else if ((i = 3) | (i > 1 & y < 20)) post(x)
           else THEN(mutate(optim(x),s,i), mutate(post(x),s,i))))
```

Finally, **crossover** is also defined structurally and is similar to mutation. A crossover method is defined for crossing integers, for crossing each of the terms, and recursively crossing their components. A term can be crossed directly with another term or one of its sub-components. The idea of crossover is to find an average, middle-point between two values. For an integer, this is straightforward. For two terms from different classes, we pick a "middle" class based on the hierarchical representation of the algebra tree. When this makes sense, we use the sub-terms that are available to fill the newly generated class. For two terms from the same class, we apply the crossover operator recursively.

The influence of genetic algorithms [Gol89][Ree93] in our learning loop is obvious, since we use a pool of terms to which we apply crossover and Darwinian selection. However, crossover is mixed with mutation that is the equivalent of parallel randomized hill-climbing. *Hill-climbing* translates the fact that we select the best terms at each iteration (cf. next section). *Randomized* tells that the neighborhood of a term is very complex and we simply randomly pick one neighbor. *Parallel* comes from the use of a pool to explore different paths in simultaneous ways. The use of a mutation level index shows that we use three neighborhood structures at the same time. We shall see in Section 5 that using a pure genetic algorithm framework turned out to be more difficult to tune, which is why we are also applying a hybrid approach at the learning level.

## 4.2 The Learning Loop

The learning process is performed with a serie of iterations and works with a pool of terms (i.e. algorithms) of size M. During each iteration, terms are evaluated and the K best ones are selected and mutated and crossed in different ways. Experimental parameters govern how many terms will be generated via mutation, crossover, and invention along with how many of the very best ones from the previous generation will be kept (K). We shall see in Section 4.2 the relative influence of these different techniques. After completing all experiments, the best terms are rated by running tests using different data sets and averaging over all runs. This second step of evaluation, described as "more thorough" in Figure 3, is explained in the following sub-section. The goal is to identify the best term from the pool in a more robust (stable) manner.

The best term is always kept in the pool. For each of the K "excellent" terms, we apply different levels of mutation (1,2 and 3), we then cross them to produce (K * (K – 1) / 2) new terms and finally we inject a few invented terms (M – 1 – K * 3 – (K * (K – 1) / 2)). The new pool replaces the old one and this process (the "learning loop") is iterated I times.

We have tried to increase the efficiency of the search by taking the distance to the best term as a parameter for choice for the other "excellent" terms, but this was not successful. The use of this distance to increase the diversity in the pool subset resulted in a lower ability to finely explore the local neighborhood. The use of a tabu list is another direction that we are currently exploring but the difficulty is to define the semantic of this list in a world of potentially infinite terms.

The number of iterations I is one of a number of parameters that need to be set for a serie of experiments. Additional parameters include the pool size M and a set of parameters governing the break down of the pool (i.e., how many terms generated via invention, mutation, and cross-over kept from the previous generation). In Section 5, we describe a set of experiments aimed at determining "good" values for each of these parameters.

To avoid getting stuck with a local minimum, we do not run the learning loop for a very large number of iterations. When we started studying the stability of the results and the influence of the number of iterations, we noticed that the improvement obtained after 20 iterations are not really significant, while the possibility of getting stuck with a local minimum of poor quality is still high. Thus we decided to divide the set of N iterations of the previous learning loop as follows (we pick N = 50 to illustrate our point).

- We compute the number R of rounds with R = N / 12 (e.g., 50 / 12 = 4)
- For each round, we run the learning loop 10 times, starting from a pool of randomly generated terms.
  We keep the best term for each round. (Note: If N is less than 24, only a single round is executed)
- We create a new pool with these k best terms, completed with random invention and apply the learning loop for the remaining number of iterations (N – R * 10).

The following figure summarizes the structure of the learning algorithm.

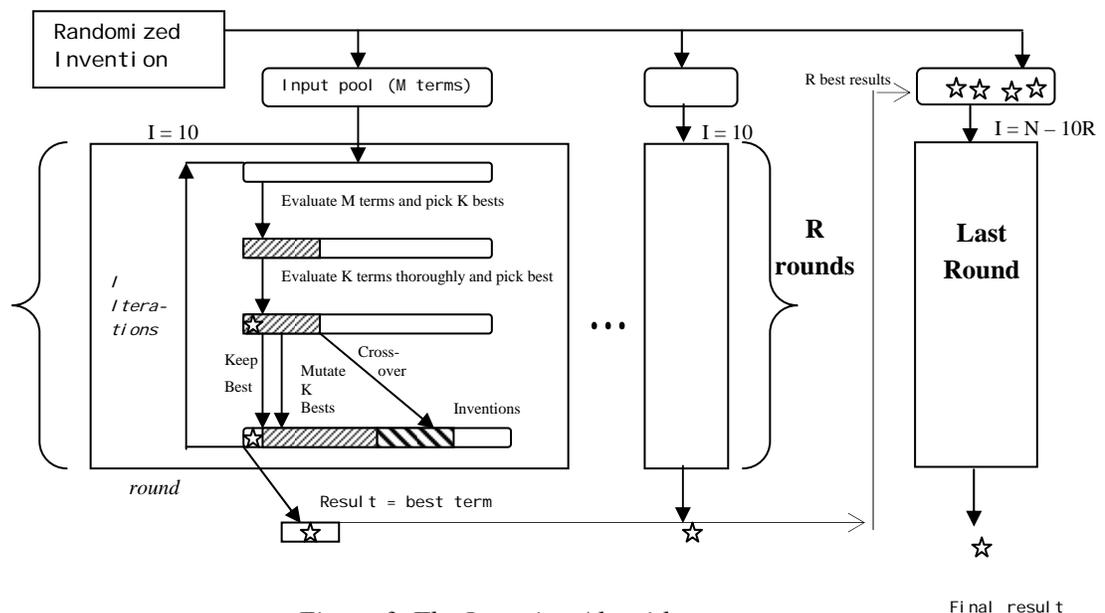

*Figure 3. The Learning Algorithm*

This figure makes it clear that the algorithm is more complex than the preliminary version that was presented in [CSL99]. The main reason for this complexity is the increase in robustness. We shall see in section 6.1 that that this more sophisticated algorithm produces better terms, but the most important improvement is the stability and statistical significance of the results that will be shown in Section 5. In particular, we discuss the importance of the training set in Section 5.4. We have used our approach in two types of situations: a machine learning paradigm where the algorithm is trained on a data set and tested on a *different* data set, and also a data mining paradigm where the algorithm is trained on the same set onto which it is applied. Section 5 will also deal with the tuning of the learning loop parameters (N, R, K, …).

## 4.3 Randomization and Further Tuning

Randomized algorithms such as LNS have a built-in instability, in the sense that different runs will produce different results for the same algorithm. Thus it is difficult to compare terms using one run only, even though the results are already averaged using a data set. A better approach is to perform multiple runs for each term, but this is expensive and slows the learning process quite a bit. The compromise that we make is that we first sort the whole pool of term using a single run to select the subset of "excellent terms", and then run multiple experiments on these selected terms to sort them more precisely. This is important to increase the probability of actually selecting the best term of the pool.

Another issue is the minimization of the standard deviation associated with a randomized algorithm. Once an algorithm (i.e., a term) incorporates a randomized technique, its performance is not simply measured by the value of one run or the average of multiple runs, but also by the standard deviation of the values during these multiple runs. In order to ensure that we minimize this deviation as well as the average value, we enrich the semantic of the evaluation of the sub-pool of "excellent" terms. When we perform multiple runs on these selected terms, the final value associated to each term is the average of the worst value and the average value. Thus, we discriminate against terms that have a high deviation.

The balancing between minimizing the average and the standard deviation is problem-dependent and the definition of a unique quality indicator only makes sense in a given industrial settings. For some problems such as on-line optimization, standard deviation is important since a poor solution translates into customer non-satisfaction. For some batch problems, the goal is purely to save money and minimizing the average value is quite sufficient. Thus, we report three numbers for each algorithm:

- The average value, **E(v)** that represents the average of the value of the term produced by the algorithm over multiple learning experiments. If we define $v$ as the valuation of the term t produced at the end of the learning loop according the definition given in Section 3.2, we have:

  $$\mathbf{E(v)} = \frac{1}{L} \times \sum_{i \leq L} v_i$$

  assuming that we ran the learning process L times and obtained the values $v_1, \ldots v_L$.

- The standard deviation $\sigma(\mathbf{v})$ represents the standard deviation of the previously defined value **v** over the multiple learning runs:

  $$\sigma(\mathbf{v}) = \sqrt{\left(\frac{1}{L} \times \sum_{i \leq L} v_i^2\right) - E(v)^2}$$

- Last, we also measure the average standard deviation **E(σ)**, since we recall that the terms produced by our learning algorithm represent randomized algorithms and that their value $v_i$ is simply an average. More precisely, to produce a value for a term, we make 10 runs of the algorithm on each test file of the data set and we record both the average value $v_i$ and the standard deviation $\sigma_i$. We may then define **E(σ)** to represent the average of this standard deviation, when multiple learning experiments are made:

  $$\mathbf{E(\sigma)} = \frac{1}{L} \times \sum_{i \leq L} \sigma_i$$

Obviously, reporting a standard deviation falls short of characterizing a distribution. However, our experience is that standard deviation is a reasonable first-order indicator for all practical purposes. For instance, the difficulties that we experienced with the early approach of [CLS99], which we report in the next section, did translate into high standard deviation and the reduction of **E(σ)** was indeed a symptom of the removal of the problem.

The three techniques that we use for creating new terms (invention, mutation and crossover) have the potential to develop terms that are overly complex. The complexity bound that guides each of these tools can prevent the creation of terms that are too expensive to evaluate. However, it is still possible to generate terms, which meet the complexity bound, but which are overly long and complex. Large terms are undesirable for two reasons: they are difficult to read and they tend to get out of hand quickly (large terms crossed with large terms generally produce even larger terms). For this reason, we have introduced a "diet" function, which takes a bound on the physical size (i.e., number of subterms and components) of the term. This function is quite simple: if the term is too large (too many sub-terms), the sub-terms are truncated at an arbitrary distance from the root (of the tree that describes the term).

## 5. Learning Characteristics

### 5.1 Convergence Rate and Stability

The algorithms defined by the terms have a number of randomness issues. As noted above, algorithms such as LNS are randomized algorithms and hence, testing a term over a single data set could yield different results. Also, the process of learning itself is randomized. To deal with these issues, we must be careful to average each of our experiments over multiple runs, recording the standard deviation over all runs, as explained in the previous section.

The learning process itself runs over a number of iterations. Showing that there is in fact a "convergence" (i.e., the stability and quality of the results should improve as the number of iterations grows) and determining the number of iterations achieves the best tradeoff between processing time and producing the best results are two key issues. To determine this we have run a serie of experiments over the same problem running from 10 iterations to 50 iterations of the learning loop. For each experiment we report the three measures: $E(v)$, $\sigma(v)$, and $E(\sigma)$, as explained earlier. We ran these experiments twice, using two different complexity goals of respectively 50K (columns 2 to 4) and 200K insertions (columns 5 to 7). The goal here is to minimize the number of trucks. The results are described in Table 2:

| Goal | $E(v)$ 50K | $\sigma(v)$ 50K | $E(\sigma)$ 50K | $E(v)$ 200K | $\sigma(v)$ 200K | $E(\sigma)$ 200K |
|---|---|---|---|---|---|---|
| **10 iterations** | 39706 | 5842 | 1092 | 34600 | 4679 | 1193 |
| **20 iterations** | 34753 | 5433 | 775 | 31303 | 4040 | 740 |
| **30 iterations** | 35020 | 4663 | 1426 | 31207 | 2090 | 755 |
| **40 iterations** | 34347 | 4623 | 1142 | 30838 | 3698 | 866 |
| **50 iterations** | 32173 | 2069 | 1217 | 29869 | 1883 | 862 |

*Table 2. Influence of the number of iterations*

A number of observations can be made from these results. We see that the quality of the terms is good, since they are clearly better than those that were hand-generated. This point will be further developed in the next section. We also see that the standard deviation of the generated algorithm is fairly small, but tend to increase when the number of iteration rises. This is due to the fact that better terms are making a heavier use of the randomized LNS. Last, we see that the standard deviation of the average value is quite high, which means that our learning algorithm is not very robust, even though it is doing a good job at finding terms.

It is interesting to notice that this aspect was even worse with the simpler version presented in [CSL99], where we did not partition the set of iterations into rounds. In that case, while the average value goes down with a higher number of iterations, the worst-case value is pretty much constant, which translates into a standard deviation that **increases** with the number of iterations.

In practice, the remedy to this instability is to run the algorithm with a much larger number of iterations, or to take the best result of many separate runs, which is what we shall do in the last section where we attempt to invent higher-quality terms.

## 5.2 Relative contribution of Mutation, crossover, and invention

The next set of experiments compares four different settings of our Learning Loop, which emphasize more heavily one of the term invention techniques. All settings use the same pool size, the difference comes from the way the pool is re-combined during each iteration:

- **al2** is our default setting, the pool size is 16 terms, out of which 3 "excellent terms" are selected, producing 9 mutated terms and 3 products of crossover, completed by 3 newly invented terms.
- **am2** is a setting that uses a sub-pool of 4 "excellent" terms, yielding 12 mutated terms, but only one crossover (the two best terms) and 2 inventions.
- **ag2** is a setting that uses the crossover more extensively, with only one mutation (of the best term), but 10 crossovers produced from the 5 "excellent" terms.
- **ai2** is a setting that uses 6 mutated terms and 1 crossover, leaving the room for 8 invented terms.

The next table reports our results using the 50K goals for term complexity and the same set of data tests. We report results when the number of iteration is respectively set to 20 (one large round) and 40 (three rounds).

| *#of iterations* | $E(v)$ 20 | $\sigma(v)$ 20 | $E(\sigma)$ 20 | $E(v)$ 40 | $\sigma(v)$ 40 | $E(\sigma)$ 40 |
|---|---|---|---|---|---|---|
| **Al2** | 34753 | 5433 | 775 | 31303 | 4040 | 740 |
| **Am2** | 34996 | 3778 | 1091 | 33832 | 2363 | 355 |
| **Ag2** | 40488 | 5627 | 1060 | 36974 | 4557 | 1281 |
| **Ai2** | 37209 | 4698 | 566 | 32341 | 2581 | 374 |

*Table 3. Comparing Mutation, Crossover and Invention*

We can see that the better approach is to use a combination of techniques (al* family), as we have selected in the rest of the experiments.

These results also suggests that mutation is the strongest technique, which means that randomized hill-climbing is better suited for our learning problem than a regular genetic algorithm approach. This result is probably due to the inadequacy of our crossover operator. It should be noticed that when we introduced crossover, it brought a significant improvement before we refined our mutation operator as explained in Section 4.1 (i.e., at that time, al* was much better than am*). We also notice that invention also works pretty well, given the time (ai2 is a blend of mutation and invention). The conclusion is that the randomization of the "evolution" technique is very important, due to the size of the search space. A logical step, which will be explored in the future, is to use a more randomized crossover.

## 5.3 Number of Iterations versus Pool Size

There are two simple ways to improve the quality of the learning algorithm: we can increase the number of iterations, or we can increase the size of the pool. The next set of experiments compare 4 settings, which are roughly equivalent in term of complexity but use different pool size vs. number-of-iterations compromise:

- **al3**, which is our regular setting (cf. al2) with 30 iterations (whereas al2 uses 20 iterations)
- **ap0**, with a smaller pool size (10), where only two best terms are picked to be mutated and crossed. The number of iteration is raised to 50.
- **ap1**, with a pool size of 35, with 6 best terms chosen for mutation and 5 for crossovers. The number of iteration is reduced to 15.
- **ap2**, with a pool size of 24, with 5 best terms chosen for mutation, and 4 for crossovers. The number of iterations is set to 20.

These experiments are made with two complexity objectives, respectively 50 000 and 200 000 insertions.

| Goal | E(v) 50K | σ(v) 50K | E(σ) 50K | E(v) 200K | σ(v) 200K | E(σ) 50K |
|---|---|---|---|---|---|---|
| Al3 | 35020 | 4663 | 1426 | 31207 | 2090 | 755 |
| Ap0 | 34745 | 3545 | 829 | 31735 | 1912 | 189 |
| Ap1 | 35101 | 4920 | 751 | 30205 | 1905 | 556 |
| Ap2 | 37122 | 4910 | 979 | 31625 | 2401 | 468 |

*Table 4. Pool Size versus Number of Iterations*

These results show that a compromise must be indeed be found, since neither using a very large pool nor using a small one seems a good idea. Our experience is that a size around 20 seems the best trade-off, but this is based on our global bound on the number of iterations, which is itself limited by the total CPU time. Many of these experiments represent more than a full day of CPU. When faster machines are available, one should re-examine the issue with a number of iterations in the hundreds.

## 5.4 Importance of training set

Finally, we need to study the importance of the training set. It is a well-known fact for researchers who develop hybrid algorithms for combinatorial optimization that a large set of benchmarks is necessary to judge the value of a new idea. For instance, in the world of job-shop scheduling, the set of ORB benchmarks is interesting because techniques that significantly improve the resolution of one problem often degrade the resolution of others. The same lesson applies here: if the data set is too small, the learning algorithm will discover "tricks" of little value since they do not apply generally. To measure the importance of this phenomenon, we have run the following experiments, using the same corpus of the 12 R1* Solomon benchmarks:

- **rc1**: train on 12 data samples, and measure on the same 12 data samples
- **rc2**: train on the first 6, but measure on the whole set
- **rc3**: train on the 12 data samples, but measure only on the last 6 (reference point)
- **rc4**: train on the first 6, measure on the last 6
- **rc5**: train on last 6 and measure on the last 6.

We use our "al2" setting for the learning algorithm, with a complexity goal of 50K insertions.

|  | **E(v)** | **σ(v)** | **E(σ)** |
|---|---|---|---|
| **rc1** | 37054 | 4154 | 1136 |
| **rc2** | 34753 | 5433 | 983 |
| **rc3** | 35543 | 3362 | 1636 |
| **rc4** | 34753 | 5433 | 1418 |
| **rc5** | 29933 | 4053 | 2015 |

*Table 5a. Impact of the training set (I)*

These results can be appreciated in different manners. On the one hand, it is clear that a large training set is better than a smaller one, but this is more a robustness issue, as shown by the degradation of the standard deviation with a smaller training set. Surprisingly, the average value is actually better with a smaller subset. We also made experiment with a training set of size 2 or 3, and the results were really not good, from which we concluded that there is a minimal size around 5. On the other hand, the sensitivity to the training set may be a desirable feature, and the degradation from learning with 6 vs. 12 data samples is not very large. This is even true in the worst case (rc4) where the term is trained from different samples than those that are used to test the result. The only result that is significantly different is rc5, which corresponds to the case where we allow the algorithm to discover "tricks" that will only work for a few problems. The gain is significant, but the algorithm is less robust. The other four results are quite similar.

The following table shows the results obtained with training and testing respectively on the R1*, C1* and RC1* data sets from the Solomon benchmarks. We added a fourth line with a term that represents out "hand-tuned" hybrid algorithm, which we use as a reference point to judge the quality of the invented terms.

| evaluation → | R1* | C1* | RC1* |
|---|---|---|---|
| train on R1* | 41555 | 98267 | 65369 |
| train on C1* | 61238 | 98266 | 79128 |
| train on RC1* | 42504 | 98266 | 66167 |
| hand-tuned term | 50442 | 98297 | 78611 |

*Table 5b. Impact of the training set (II)*

If we replace the difference into perspective using the reference algorithm, we see that the learning algorithm is not overly sensitive to the training set. With the exception of the C1* data set, which contains very specific problems with clusters, the algorithms found by using R1* and RC1* do respectively well on the other data set. In the experiments reported previously in this section, we have used the same set for training and evaluating. For industrial problems, for which we have hundreds of sample, we can afford to use separate sets for training and evaluation, but the results confirm that the differences between a proper machine learning setting (different sets) and a "data mining" setting (same set) are not significant.

Our conclusion is that we find the learning algorithm to be reasonably robust with respect to the training set, and the ability to exploit specific features of the problem is precisely what makes this approach interesting for an industrial application. For instance, a routing algorithm for garbage pick-up may work on problems that have a specific type of graph, and for which some meta-heuristic combinations are well suited, while they are less relevant for the general case. The goal of this work is precisely to take this specificity into account.

# 6. Learning Experiments

## 6.1 Hand-crafted vs. discovered terms

The first experiment that we can make is to check whether the terms that are produced by the learning algorithm are indeed better than we could produce directly using the algebra as a component box. Here we try to build a term with a complexity of approximately 50Ki (which translates into a run-time of 2s on a PIII-500MHz), which minimizes the total travel time. The data set is the whole R1* set. In the following table, we compare:

- four terms that were hand-generated in an honest effort to produce a good contender.
- the term *succLNS* that we have shown in Figure 1 to be the best among our introduction examples, which tries to emulate the LNS strategy of [Sha98].
- the term (Ti1) that we presented in [CSL99], that is the output of our first version of the learning algorithm.
- the term (Ti2) that is obtained with the new learning algorithm with the **al2** settings and 40 iterations.

| *Term* | *Objective* | *Run-time (i)* |
|---|---|---|
| LDS(3,5,0) | 22669 | 1.3Ki |
| DO(LDS(3,3,100), LOOP(8,LNS(10,4,LDS(4,4,1000)))) | 22080 | 99Ki |
| DO(LDS(3,2,100),TREE (20,2)) | 22454 | 23Ki |
| FORALL(LDS(3,2,100),CHAIN(8,2)) | 22310 | 59Ki |
| **SuccLNS (cf. Table 1)** | 21934 | 59Ki |
| Ti1:     FORALL(LDS+(0,2,2936,CHAIN(1,1)), LOOP(48,LOOP(4, LNS(3,16,LDS(3,1,287))))) | 21951 (invented [CSL99]) | 57ki |
| Ti2: DO(INSERT(0), THEN(LOOP(22,THEN( LOOP(26,LNS(6,22,LDS(3,1,196))), LNS(5,23,LDS(2,1,207)))), LNS(4,26,LDS(1,1,209)))) | 21880 (invented !) | 57ki |

*Table 6. Inventing new terms (travel minimization)*

These results are quite good, since not only the newly invented term is clearly better than the hand-generated example, but it is even better than the complex *SuccLNS* term that is the implementation with our algebra of a strategy found in [Sha98], which is itself the result of careful tuning.

The term shown in the table is the *best* term found in 5 runs. The *average* value for term Ti2 was 21980, which is still much better than the hand-generated terms. It is also interesting to note that the results of the algorithms have been improved significantly by the tuning of the mutation operator and the introduction of a new control strategy for the Learning Loop, as explained in Section 3.

## 6.2 Changing the Objective function

We now report a different set of experiments where we have changed the objective function. The goal is now to minimize the number of trucks (and then to minimize travel for a given number of trucks). We compare the same set of terms, together with two new terms that are invented using this different objective function:

- the term (Ti3) that we presented in [CSL99], that is the output of our first version of the learning algorithm.

- the term (Ti4) that is obtained with the new learning algorithm with the **al2** settings and 40 iterations.

| Term | Objective | Run-time (i) |
|---|---|---|
| LDS(3,5,0) | 68872 | 1.3Ki |
| DO(LDS(3,3,100), LOOP(8,LNS(10,4,LDS(4,4,1000)))) | 47000 | 99Ki |
| DO(LDS(3,2,100),TREE(20,2)) | 47098 | 23Ki |
| FORALL(LDS(3,2,100),CHAIN(8,2)) | 54837 | 59Ki |
| **SuccLNS (cf. Table 1)** | 40603 | 59Ki |
| **Ti2 (cf. Table 5)** | 525000 | 59Ki |
| Ti3:DO(LDS(2,5,145), THEN(CHAIN(5,2),TREE(9,2,2))) | 36531 (invented [CSL99]) | 57ki |
| Ti4:DO(LDS(2,4,178), THEN(LOOP(3,LOOP(81,LNS(7,2,LDS(2,1,534)))), LNS(3,7,LDS(2,1,534))), LOOP(3,LOOP(83,LNS(4,3,LDS(2,1,534)))), LNS(3,25,LDS(0,2,983))))) | 28006 (invented !) | 57ki |

*Table 7. Inventing new terms (travel minimization)*

These results are even more interesting since then invented term is much better than the other one, showing a very significant improvement since [CSL99]. The fact that the strategy used to build SuccLNS was proposed to minimize the number of trucks makes the difference between the terms quite surprising. The average number of routes corresponding to Ti4 is 12.33, which is excellent, as we shall discuss in the next section. The term Ti4 is the best found in a set of 5 experiments and is not totally representative since the average value was found 32600, which is still much better than the hand-generated terms.

It is interesting to notice that the term Ti3 is not good either as far as travel optimization is concerned (the value of the objective function accentuates strongly the decrease in the quality of the solution, because the value of an ideal solution is subtracted from it). The specialization of the term according to the objective function is thus proven to be an important feature. This is even more so in an industrial application context, where the objective function also includes the satisfaction of soft constraints such as operator preferences. These constraints are business-dependent and they change rapidly over the years.

## 6.3 Changing the complexity goal

Here we report the results found with different complexity goals, ranging from 50K to 1200K insertions. These experiments were made on a PIII-800Mhz machine.

| | Value | Equivalent # of trucks | Complexity | Equivalent run-time |
|---|---|---|---|---|
| 50K | 28006 | 12.33 | 56K | 2s |
| 200K | 29732 | 12.44 | 211K | 10s |
| 600K | 26000 | 12.26 | 460K | 20s |
| 1200K | 24459 | 12.25 | 1200K | 40s |

*Table 8. Inventing new terms*

These results are impressive, especially with a limited amount of CPU time. In a previous paper, using the best combination of LNS and ILO and restricting ourselves to 5 minutes of CPU time, we obtained an average number of route of 12.31 on the R1* set of Solomon benchmarks, whereas [TGB95] and [Sha95] respectively got 12.64 and 12.45, using 10 minutes for [Sha98] or 30 minutes for [TGB95] (the adjustment in CPU time is due to the difference in hardware). Here we obtain a much better solution (12.25 routes on R1*) in less than 1 minutes of CPU time. This is even better that was previously obtained when we augmented the available CPU time by a factor of 3: 12.39 [Sha98] and 12.33[TGB95]. This clearly shows that the algorithms that were found by our learning methods produce state-of-the-art results.

The improvement with respect to [CSL99] is also obtained on the RC1* set, since the term obtained with complexity 600K has an average number of routes of 11.875 in 30s of CPU time, whereas our best strategy in [CSL99] obtained 12.0 routes in 5 minutest and whereas [TGB95] and [Sha95] respectively got 12.08 and 12.05.

## 6.4 Future Directions

This paper has shown that this learning method produces precise tuning for hybrid algorithms applied to medium-sized problems and short to medium run-times (seconds or minutes). Depending on the parameters and the size of the data set, the ratio between the learning loop and the final algorithm run-times varies from 1000 to 10000, with a practical value of at least 10000 to get stable and meaning-full results. For larger problems for which a longer run-time is expected, this is a limiting factor. For instance, we use a hybrid algorithm for the ROADEF challenge [Roa01] that can indeed be described with our algebra. However, the target run-time proposed in the challenge is one hour, which translates into more than a year for the learning process. Needless to say, we could not apply this approach and used manual tuning that produced very competitive results [Roa01].

Thus, we either need to find a way to train on smaller data set and running time, or to shorten the training cycle. The first direction is difficult; we found that going from medium-sized VRP to larger one seemed to preserve the relative ordering of hybrid algorithms, but this is not true of frequency assignment problems given in the ROADEF challenge. On the other hand, using techniques found for smaller problems is still a good idea, even though parameters need to be re-tuned. Thus we plan to focus on "warm start" loops, trying to cut the number of cycles by feeding the loop with terms that are better than random inventions.

Our second direction for future work is the extension of the algebra. There are two operators that are natural to add in our framework. First, we may add a MIN($n,t$) operator that applies $n$ times a build-or-optimize term $t$ and selects the best iteration. This is similar to LOOP($n,t$) except that LOOP applied $t$ recursively to the current solution, whereas MIN applies it $n$ times "in parallel". The other extension is the introduction of a simple form of tabu search [Glo86] as one of the meta-heuristics in the "tool box", which could be used instead of LNS. The interest of a tabu search is to explore a large number of simple moves, as opposed to LNS, which explores a small number of complex moves. In our framework, the natural tabu exploration is to apply push and pull as unit moves.

# 7. Conclusion

We presented an approach for learning hybrid algorithms that can be expressed as a combination of meta-heuristics. Although these experiments were only applied to the field of Vehicle Routing with Time Windows, hybrid algorithms have shown very strong results in many fields of practical combinatorial optimization. Thus, we believe that the technique that we propose here has a wide range of applicability. We showed how to represent a hybrid program with a term from an algebra and presented a learning algorithm that discovers good algorithms that can be expressed with this algebra. We showed that this learning algorithm is stable and robust, and produces terms that are better than those that a user would produce through manual tuning.

The idea of learning new optimization algorithm is not new. One of the most interesting approaches is the work of S. Minton [Min97], which has inspired our own work. However, we differ through the use of a rich program algebra, which is itself inspired from the work on SALSA[LC98], and which enables to go much further in the scope of the invention. The result is that we can create terms that truly represent state-of-the-art algorithms in a very competitive field. The invention in [Min97] is mostly restricted to the choice of a few heuristics and control parameters in a program template. Although it was already shown that learning is already doing a better job than hand tuning in such a case, we think that the application of learning to a hybrid program algebra is a major contribution to this field. The major contribution of this paper is to provide a robust and stable algorithm, which happens to be significantly better than our preliminary work in [CSL99].

This work on automated learning of hybrid algorithm is significant because we found that the combination of heuristics is a difficult task. This may not be a problem to solve well-established hard problems such as benchmarks, but it makes using hybrid algorithms for industrial application less appealing. The hardware on which these algorithms are running is changed every 2 or 3 years with machines that are twice faster; the objective function of the optimization task changes each time a new soft constraint is being added, which happens on a continuous basis. For many businesses, the size of the problems is also evolving rapidly as they need to service more customers. All these observations means that the optimization algorithm needs to be maintained regularly by experienced developers, or it will become rapidly quite far from the state-of-the-art (which is what happens most of the times, according to our own industrial experience). Our goal is to incorporate our learning algorithm to optimization applications as a self-adjusting feature.

To achieve this ambitious goal, more work is still needed. First, we need to demonstrate the true applicability of this work with other domains. We are planning to evaluate our approach on another (large) routing problem and on a complex bin-packing problem. We also need to continue working on the learning algorithm to make sure that it is not wasting its computing time and that similar or even better terms could not be found with a different approach. Another promising future direction is the use of a library of terms and/or relational clichés. By collecting the best terms from learning and using them as a library of terms and/or deriving a set of term abstractions like the clichés of [SP91], it will be possible to bootstrap the learning algorithm with a tool box of things that have worked in the past.

## Acknowledgments

The authors are thankful to Jon Kettenring for pointing out the importance of a statistical analysis on the learning process. Although this task is far from complete, many insights have been gain through the set of systematic experiments that are presented here.

The authors are equally thankful to the anonymous reviewers who provided numerous thought-provoking remarks and helped us to enrich considerably the content of this paper.